\documentclass[aps,bibnotes,twocolumn,final,floatfix,balancelastpage]{revtex4-1}

\usepackage{amsmath}
\usepackage[utf8]{inputenc}
\usepackage{mathptmx}
\usepackage[T1]{fontenc}
\usepackage{hyperref}
\hypersetup{colorlinks=true,linkcolor=blue,citecolor=blue,urlcolor=blue,filecolor=blue}

\usepackage{graphicx}
\usepackage{tikz}
\usepackage{microtype}
\usepackage{xspace}

\newlength\imagewidth
\newlength\imagescale

\newcommand{\be}{\begin{equation}}
\newcommand{\ee}{\end{equation}}
\newcommand{\quant}[2]{$#1\,\text{#2}$}

\newcommand{\compositionformula}[4]{(#2)$_{#1}$(#4)$_{#3}$}
\newcommand{\boratecompositionformula}[3]{\compositionformula{#1}{#2$_2$O}{#3}{B$_2$O$_3$}}
\newcommand{\lowrubidium}{\boratecompositionformula{2}{Rb}{98}\xspace}
\newcommand{\highrubidium}{\boratecompositionformula{15}{Rb}{85}\xspace}
\newcommand{\higherrubidium}{\boratecompositionformula{30}{Rb}{70}\xspace}
\newcommand{\highcaesium}{\boratecompositionformula{15}{Cs}{85}\xspace}
\newcommand{\Rbformel}{\boratecompositionformula{x}{Rb}{(1-x)}\xspace}
\newcommand{\Csformel}{\boratecompositionformula{x}{Cs}{(1-x)}\xspace}

\begin{document}

\title{Beam-induced Atomic Motion in Alkali Borate Glasses}

\author{Katharina Holzweber}
\email{katharina.holzweber@univie.ac.at}
\affiliation{Fakult\"at f\"ur Physik, Universit\"at Wien, Boltzmanngasse 5, 1090 Wien, Austria}
\author{Christoph Tietz}
\affiliation{Fakult\"at f\"ur Physik, Universit\"at Wien, Boltzmanngasse 5, 1090 Wien, Austria}
\author{Tobias Michael Fritz}
\affiliation{Fakult\"at f\"ur Physik, Universit\"at Wien, Boltzmanngasse 5, 1090 Wien, Austria}
\author{Michael Leitner}
\affiliation{Heinz Maier-Leibnitz Zentrum (MLZ), Technische Universit\"at M\"unchen, Lichtenbergstra\ss e 1, 85747 Garching, Germany}
\author{Bogdan Sepiol}
\affiliation{Fakult\"at f\"ur Physik, Universit\"at Wien, Boltzmanngasse 5, 1090 Wien, Austria}

\date{\today}

\begin{abstract}
Applying coherent X-rays by the method of atomic-scale X-ray Photon Correlation Spectroscopy results in beam-induced dynamics in a number of oxide glasses. Here these studies are extended to rubidium and caesium borates with varying alkali contents. While no cumulative beam damage is observed, the observed rate of structural rearrangements shows a linear relation to the dose rate. In agreement with the increasing glass transition temperature, the rate of dynamics at given dose rate decreases with increasing alkali content, while the shape of the decay of correlations becomes progressively stretched. This behavior is explained in terms of faster dynamics of the alkali positions compared to the borate network. Finally, the $q$-dependent behavior of the correlation decay rate implies the observed dynamics to proceed via small-scale atomic displacements subject to de Gennes narrowing.
\end{abstract}

\maketitle

\section{Introduction}
\label{sec:intro}
Examining the mechanisms of diffusion in solids on an atomic level is a fundamental issue. Decisive experimental insights have been obtained by the application of classical methods such as Quasi-Elastic Neutron Scattering (QENS) or Mössbauer Spectroscopy~\cite{vogldiffcondmat2005}. However, the limitation of these methods to comparatively fast diffusivities and specific elements or isotopes has confined such direct investigations to favorable prototypical systems. X-ray Photon Correlation Spectroscopy (XPCS) holds the promise to fill this gap, as it deals with coherent scattering in the time domain. While in the pioneering papers by Sutton et al.~\cite{sutton1991} and Brauer et al.~\cite{Brauer1995} metallic systems were studied, the majority of later studies concerned soft matter dynamics, with objects in the nanometer range (see e.g. Refs.~\cite{Robert2005, gruebel2008, cipelletti2011, chushkin2012, Hruszkewycz2012, guo2012}). The full potential of the method has been realized by the first demonstration of atomic-scale XPCS (aXPCS) by Leitner et al.~\cite{Leitner2009}. Since then it has been applied in several crystalline~\cite{stana2014atomic} and amorphous solids \cite{ross2014direct,ruta2012atomic,giordano2016unveiling}. 

For soft condensed matter such as biological samples, the issue of potential beam damage has always been recognized, since XPCS measurements are coupled to powerful synchrotron sources with a highly intense X-ray beam. A frequent sample replacement~\cite{fluerasu2008x}, changing the exposure spot on the sample, or introducing Low Dose X-ray Speckle Visibility Spectroscopy~\cite{verwohlt2018low} are ways of mitigating this problem. The influence of the X-ray beam on hard condensed matter has been, however, more or less neglected and rarely mentioned so far~\cite {leitner2015acceleration}.

A recent remarkable observation during aXPCS measurements on silicate and germanate glasses was reported by Ruta et al.~\cite{ruta2017hard}: In these systems, there is a clear dependence of the atomic motion on the incident X-ray flux, which implies that the observed dynamics is driven by the absorbed X-ray intensity. Still, no significant structural modification of the sample is evident. This is in stark difference to studies on crystalline alloys or metallic glasses, where such an effect is not seen~\cite {leitner2015acceleration, ruta2017hard}. 

The structure and dynamics of alkali borate glasses, prototypical solid-state ionic conductors and thus of great technological interest, are still topics of current research~\cite[e.g.][]{wright2010borate}. While the accessibility of beam-driven dynamics in the borate end-member system has been established in a recent publication~\cite{pintori2019relaxation}, it is not yet known how the addition of alkali oxides, which leads to well-established anomalies in a number of static and dynamical properties, collectively known as the borate anomaly~\cite{shelby2005introduction}, affects the susceptibility to a beam-driven acceleration of dynamics, much less how the rearrangements actually proceed on a microscopic scale.

Here we present a detailed study of beam-induced dynamics by aXPCS in rubidium and caesium borate glasses of different alkali contents, suppported by measurements of the glass transition temperature and Wide-Angle X-ray Scattering (WAXS). Varying the incident intensity by placing attenuators in the beam, we verify the expected linear relation of incident beam intensity with the resulting dynamics and identify the absorbed dose rate as the fundamental deciding quantity. Comparing timescales and shape parameters of the intensity fluctuation correlation functions for isoelectronic Rb- and Cs-based systems, we can distinguish the dynamics of the alkali sites from those of the borate network, while studying the correlation decay rate as function of $q$ allows us to conclude that the structural rearrangements are dominated by small-scale atomic displacements as opposed to discrete jumps on length-scales comparable to atomic distances. 

\section{Intensity autocorrelation: theory}
\label{sec:Intensity autocorrelation}
Introductions to the theory and practice of XPCS can be found in Refs.~\cite{sutton2008, abernathy1998, leitner2012studying}. The basic idea is to quantify the timescale of temporal fluctuations of the electron density on the atomic scale as probed by coherent X-ray scattering via the intensity autocorrelation function
\be
g^{(2)}(\vec{q},\Delta t)=\frac{\langle I(\vec{q},t)I(\vec{q},t+\Delta t)\rangle}{\langle I(\vec{q},t) \rangle^2}.
\label{eq:autocorrelation}
\ee
Here $I(\vec{q},t)$ is the observed intensity at scattering vector $\vec{q}$ and at time $t$ and the brackets $\langle ...\rangle$ denote the ensemble average, which experimentally is realized as an average over absolute time $t$ as well as over the corresponding pixels on the detector. This quantity can be related to the normalized intermediate scattering function $F(\vec{q}, t)$
\be
g^{(2)}(\vec{q},\Delta t)=1+\beta\bigl(F(\vec{q},{t})\bigr)^2,
\label{eq:g2_f(q,t)}
\ee
where $\beta$ is the coherence factor with values between 0 and 1 quantifying the degree of coherence. Via a Fourier-transformation in space $F(\vec{q},t)$ is related to the van Hove pair correlation function $G(\vec{r},t)$ that enables information about the dynamics of particles. 

Dynamical rearrangements in the sample lead to fluctuating scattered intensities and thus to a decay of the autocorrelation function. Empirically, the expression
\be
F(\vec{q},\Delta t)=\exp\bigl(-\Gamma(\vec{q})\Delta t\bigr)^\alpha.
\label{eq:KWW}
\ee
known as Kohlrausch-Williams-Watts (KWW) function has been found to satisfactorily describe diverse systems~\cite{madsen2010beyond}. A system of non-interacting identical random walkers, or also a concentrated non-interacting lattice gas with site exclusion would give an ideal exponential decay with shape parameter $\alpha=1$, while the different local environments in a disordered glass let us expect dynamics on varying timescales, leading to a stretched exponential decay with $\alpha<1$. Still, the primary fit parameter is the decay constant $\Gamma (\vec{{q}})$, whose variation with $\vec{q}$ encodes the spatial scale of structural rearrangements. From now on $\vec{q}$ will be replaced by $q$ due to the isotropy of glasses.
\begin{table}
  \centering
\begin{tabular*}{8cm}{@{\extracolsep{\fill}}cccccc}
x 					&0.02&0.10&0.15&0.20&0.30\\
\hline
$\rho$ (g/cm$^3$)				&1.98&2.22&2.39&2.50&2.84\\
$q_\text{max}$ (\AA$^{-1}$)	&1.62&1.70&1.81&1.82&1.86\\
$1/\mu$ ($\mu$m)			&1646&642&462&369&255\\
\end{tabular*}
  \caption{Density $\rho$, position of first diffraction peak $q_{\text{max}}$, and inverse absorption length $\mu$ at 13 keV of \Rbformel.}
  \label{Rb_properties}
\end{table}

\begin{table}
  \centering
\begin{tabular*}{8cm}{@{\extracolsep{\fill}}cccccc}
x 						&0.00&0.02&0.10&0.15&0.20\\
\hline
$\rho$ (g/cm$^3$)				&1.81&1.92&2.36&2.54&2.75\\
$q_\text{max}$ (\AA$^{-1}$)	&1.59&1.70&1.75&1.76&1.77\\
$1/\mu$ ($\mu$m)			&2884&634&158&154&108\\
\end{tabular*}
  \caption{Density $\rho$, position of first diffraction peak $q_\text{max}$, and inverse absorption length $\mu$ at 13 keV of pure B$_2$O$_3$ and \Csformel.}
  \label{Cs_properties}
\end{table}

\section{Experimental approach}
\subsection{Sample preparation}
Rubidium and caesium borate glasses with different compositions \boratecompositionformula{x}{A}{1-x} have been prepared with molar fractions of $x=0$, $0.02$, $0.10$, $0.15$, $0.20$ and $0.30$. The chemically pure materials were mixed and melted in alumina crucibles in an electrically heated muffle furnace at a temperature of \quant{1273}{K} for about $3$~hours. To avoid inhomogeneities in the glass the melt was stirred several times and subsequently poured into a cylindrical brass mould. After the preparation the glasses were heat-treated at a temperature of about \quant{20}{K} below $T_{\text{g}}$ for $3$~hours and then slowly cooled down to room temperature with cooling rates on the order of \quant{1}{K/min} in order to reduce tension in the vitreous materials. As alkali borates are hygroscopic, they have been kept in vacuum during the measurement and stored in a dry atmosphere at all other times.

\begin{figure}
\centering
\includegraphics{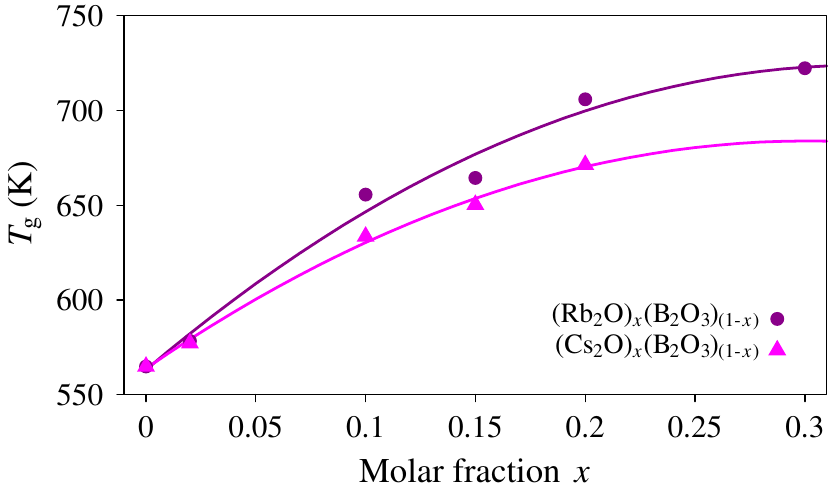}
\caption{Glass transition temperatures $T_g$ for different concentrations of \Rbformel and \Csformel.}
\label{fig:Glass_transition}
\end{figure}

The densities were measured with the method of Archimedes using decahydronaphthalene as medium and are given in Tabs.~\ref{Rb_properties} and \ref{Cs_properties}. The glass transition temperatures were determined by a Netzsch~DSC~204 device employing a heating rate of \quant{20}{K/min}. To erase any effects of the previous thermal history, we first cycled through the glass transition and determined the inflection points only in the second heating cycle. These values corresponding to the composition-dependent glass transition temperatures are given in Fig.~\ref{fig:Glass_transition}.

The actual specimens for the aXPCS measurements were prepared in the form of thin slices using a low-speed diamond saw. After grinding and polishing them a thickness of approximately 0.2 mm was obtained. Then, using a dimpling grinder, a hole in the shape of a spherical segment was excavated in order to allow choosing the sample thickness during the experiment by translating the sample table laterally. The radius of the grinding wheel was 7.5 mm. With respect to the thickness of the slices and the spot size of the beam there was almost no deviation from a plan-parallel sample in the X-ray beam. Note that for these systems, the penetration lengths at \quant{13}{keV} as used at the aXPCS measurements shown in Tabs.~\ref{Rb_properties}~and~\ref{Cs_properties} are much higher than the longitudinal coherence length of the beam. Thus, the optimal sample thickness results rather from the requirement of achieving a reasonable contrast for coherent X-ray measurements than from maximizing the scattered radiation. 

\begin{figure*}
  \includegraphics{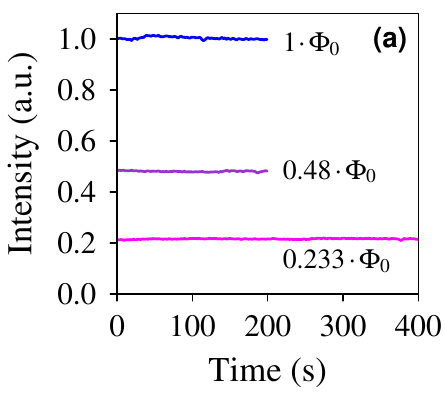}\hspace{-.5cm}
  \includegraphics{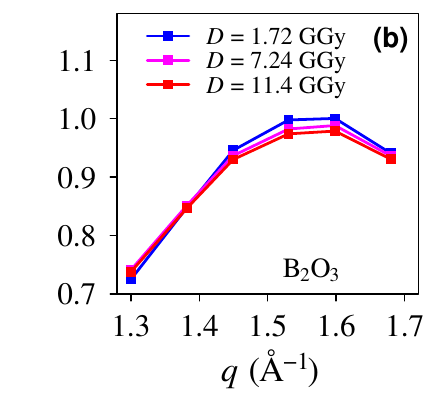}
  \includegraphics{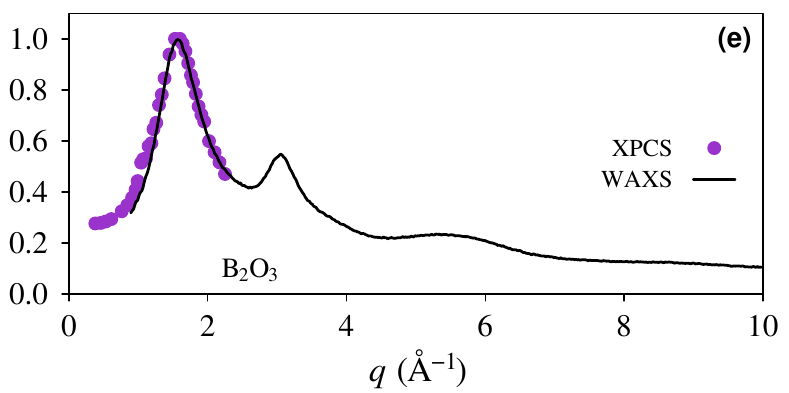}\\
  \includegraphics{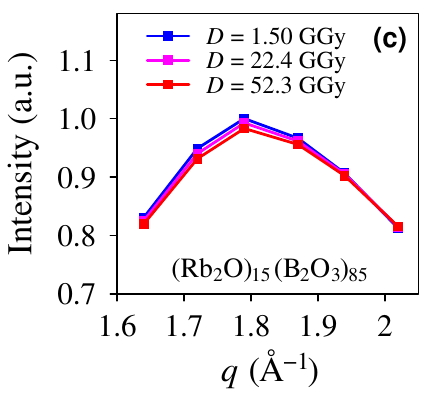}\hspace{-.5cm}
  \includegraphics{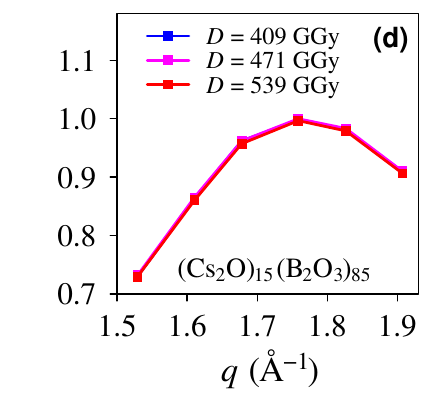}
  \includegraphics{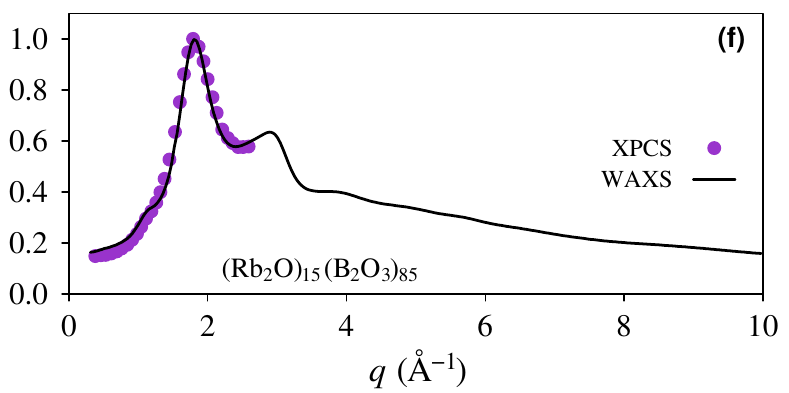}
  \caption{Absence of permanent beam damage under irradiation: (a) shows the time evolution of the intensity $I(t)$ for \highrubidium measured under different incoming $\Phi_\text{in}$ at $q_\text{max}$ with an exposure time of \quant{0.2}{s} per frame. (b)-(c) show $I(q)$ around $q_\text{max}$ measured under the same incoming $\Phi_\text{in}=\Phi_{0}$ at large time steps. (e) and (f): Comparison of WAXS-data with in-situ measured $I(q)$ for B$_2$O$_3$ and \highrubidium at room temperature. The WAXS-measurement for B$_2$O$_3$ was carried out at an Empyrean PANalytical X-ray diffractometer and for \highrubidium at beamline P02 at DESY.}
 \label{fig:Sq_static}
\end{figure*}

\subsection{WAXS measurements}
\label{sec:totalscatt}
WAXS-measurements were performed at beamline P02 of the synchrotron PETRA III at an energy of \quant{60}{keV}, as well as on an Empyrean PANalytical X-ray diffractometer at the Technical University of Vienna. The X-ray diffractometer offered no energy-dispersive detector, hence, a Cu-anode with \quant{8}{keV} was applied to avoid fluorescence. The evaluation program PDFgetX2~\cite{qiu2004pdfgetx2} was applied for obtaining the static structure factor $S(q)$. The inherent isotropy of glass systems allowed us to prepare the samples as thin slices with thicknesses around the optimal scattering lengths, thus avoiding inevitable surface crystallization and water uptake of the hygroscopic samples under the standard powder in capillary protocol.

\subsection{aXPCS measurements}
\label{sec:xpcs_measurement}
The aXPCS measurements were conducted at beamline P10 of the synchrotron PETRA III using a coherent setup with \quant{13}{keV} photons and an estimated flux of $\Phi_0=10^{11}$~photons/s. For studying the dynamics under different photon fluxes, variable numbers of attenuator plates made of Ag or Si were introduced into the beam, with thicknesses per plate of \quant{12.5}{$\mu$m} and \quant{25}{$\mu$m}, respectively, corresponding to transmission coefficients at \quant{13}{keV} of $T_\text{Ag}=0.483$ and $T_\text{Si}=0.917$ per plate. Using a refractive X-ray lenses system, a beam spot of FWHM $2.5\times2$~$\mu$m$^2$ (h$\times$v) was achieved. The sample was mounted in a custom-built chamber in transmission geometry at a vacuum of $\approx$ $10^{-6}$~mbar. All experiments were performed at room temperature. Scattered photons were recorded by the Eiger X 4M detector with a pixel size of $75\times 75\,\mu$m$^2$, a continuous readout time and a negligibly low dead time. To suppress fluorescence an appropriate energy threshold was chosen. With a sample-detector distance of about \quant{181}{cm} and a detector width of about \quant{14.5}{cm} (the actual side length of this detector is higher; but approximately 1/3 of area was not covered by the flight tube) an angle of about $4.5^{\circ}$ was covered. For higher accuracy in angular resolution we have divided the area of the detector in our evaluation program into two halves, which still provides enough coherent scattering information for reasonable statistics.

The intensity autocorrelation function was measured for several incoming fluxes $\Phi_\text{in}$ and scattering vectors $2 \theta$. Temperature- and dose-rate-dependent decay constants $\Gamma$ and shape parameters $\alpha$ were obtained by fitting Eq. \ref{eq:KWW}. A two-time correlation function $C(t_1, t_2)$~\cite{madsen2010beyond} was calculated for each measurement to check the stability of the sample and the mechanical stability of the setup.

\section{Results and discussion}
\subsection{Beam-induced effect on structural properties}
\label{sec:Structural properties}
To examine the beam-induced effect on the structure, we evaluated the intensity versus time $I(t)$ at the maximum peak of $I(q_\text{max})$. The $q_\text{max}$-values for each sample can be found in Tabs.~\ref{Rb_properties} and \ref{Cs_properties}. As Fig.~\ref{fig:Sq_static}(a) shows, there are no discernible continuous or abrupt changes in intensity while collecting a series of frames for the autocorrelation function in Eq.~\eqref{eq:autocorrelation}. This is valid for the first measurement (in Fig.~\ref{fig:Sq_static}(a) shown for \highrubidium with an incoming flux of $0.48\Phi_0$) as well as for all further measurements with different incoming fluxes, where we retained the target position.
 
For a quantitative discussion of irradiation effects, it is of course the dose rate rather than the incident flux that is relevant. Modelling the beam profile as an anisotropic Gaussian function with $d_1$ and $d_2$ as the principal full widths at half maximum gives a radiative flux density of $2 \log(2)\Phi_0/\pi d_1 d_2$ averaged over the profile of the beam, which evaluates to about \mbox{$8.825\cdot 10^{21}$} photons per second and m$^2$ for the above-mentioned parameters at full beam. Multiplying by the respective absorption cross sections, this corresponds to dose rates $S$ of a few eV per atom and second, or some ten MGy/s. 

For mapping the curvature of the first maximum peak around $I(q_\text{max})$ the area of the detector was divided into 6 equal parts, each with a $2 \theta$-range of~$\approx 0.7^{\circ}$. Figs.~\ref{fig:Sq_static}b-d shows that the shape of $I(q)$ indeed flattens slightly. Given that at these doses typically already more than one \quant{13}{keV} photon has been absorbed per atom on average and thus no further effect is expected even at much higher doses, the curvature flattening pointing to structural changes is however very small.

Combining such data for different detector positions gives us a handle on the structure of the sample during the aXPCS measurements, which can be compared with dedicated ex-situ WAXS data. As examples $I(q)$ of B$_2$O$_3$ and \highrubidium are shown in Fig.~\ref{fig:Sq_static}(e) and (f). A very satisfactory agreement is evident, which again allows us to rule out significant sample damage during our experiments, as the beam size for the WAXS measurements was much larger than in aXPCS and hence no damaging effects are expected in WAXS. This observation is in agreement with the work of Pintori et al.~\cite{pintori2019relaxation}.

\begin{figure}
\centering
\includegraphics{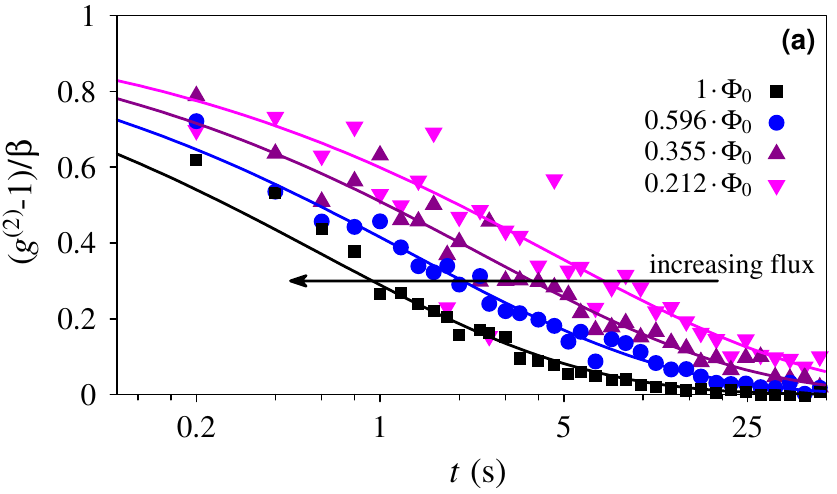}\\
\includegraphics{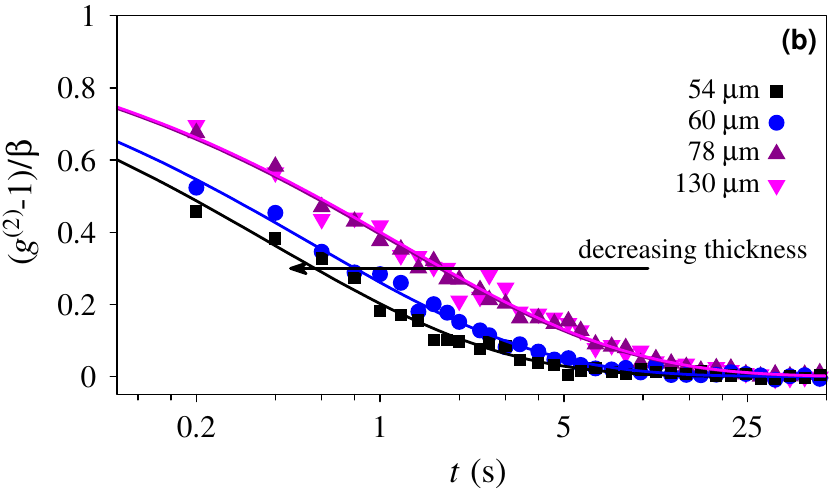}
\caption{Beam-driven dynamics in \higherrubidium: The normalized intensity autocorrelation functions measured at $q_\text{max}$ shift towards larger times for decreasing incident flux $\Phi_\text{in}=T_x\Phi_0$ with the transmission coefficient $T_x$ (a) as well as increasing thickness (b). Lines are fits with a KWW function with a common shape factor $\alpha$ in (a), and thickness-dependent $\alpha$ in (b).}
\label{fig:g2}
\end{figure}

\subsection{Beam-induced effect on dynamical properties}
\label{sec:dynamical properties}
Alkali borate glasses are described by the borate anomaly~\cite{shelby2005introduction}: The charge diffusion~\cite{berkemeier2005molar} as well as the glass transition temperature $T_g$ (Fig.~\ref{fig:Glass_transition}) increase with increasing alkali content and reach a shallow maximum at about $x\approx 0.3$ of alkali mole fraction. From a microscopic point of view this is predominately achieved by the modification of the borate network from BO$_3$ to BO$_4$ structural units~\cite{wright2010borate,verhoef1995structure}. Specifically, in pure B$_2$O$_3$ the boron atoms are in symmetric three-fold oxygen coordination. The incorporation of A$_2$O-units supply additional oxygen atoms, causing a partial increase to four-fold coordination. Four instead of only three bonds of the tetrahedrons could be a simple reason of the increasing rigidity of the system as evidenced by the rise of $T_\text{g}$. The positively charged ions preferentially reside next to the vicinity of negatively charged BO$_4$-tetrahedrons and in this constellation exhibit a very high mobility within the borate network. Beyond an alkali concentration of $x=0.3$ the ions are stronger bound to the emerging non-bridging oxygens of BO$_3$ entities that gradually replace the tetrahedrons. Hence, the ionic mobility, as well as the density and $T_g$ reverse their trend. 

Based on low charge diffusion values, e.g. $\approx 10^{-29}$(m$^2$/s) for \highrubidium at room temperature~\cite{berkemeier2005molar}, no thermal structural rearrangements visible by aXPCS can be expected for most of our systems. However, Pintori et al.~\cite{pintori2019relaxation} have demonstrated that the X-ray beam used to probe dynamics in aXPCS does give rise to measureable dynamics in pure borate glass, and that the effect is more pronounced for stronger irradiations. Fig.~\ref{fig:g2}(a) shows that this effect can indeed be reproduced in \higherrubidium: the decay of the normalized intensity autocorrelation function $g_\text{2}(q,t)$ measured at the first maximum peak of $I(q_\text{max})$ shifts towards longer times when attenuating the incident flux, revealing a slowing of dynamics with decreasing dose rate as expected for irradiation-driven dynamics.

Performing series of measurements on various target positions with different thicknesses also show a change in the decay rate of $g^{(2)}(q,t)$. Specifically, Fig. \ref{fig:g2}(b) shows that the decay rate decreases with increasing thickness. This becomes obvious considering Beer-Lambert law $\Phi(x)=\Phi_\text{in}e^{-\mu x}$, implying that also the rate of absorbed photons decline exponentially when traversing the sample.

\begin{figure}
\centering
\includegraphics{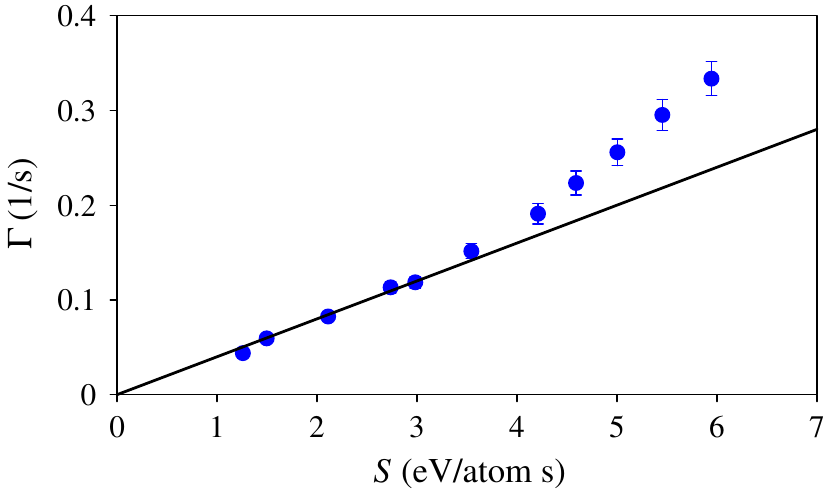}
\caption{$\Gamma$ as a function of average dose rate $S$ in \higherrubidium measured at $q_\text{max}$. The blue full circles indicate a measurement series of different combinations of absorbers, while the line corresponds to a strictly linear proportionality.}
\label{fig:30Rb_flux}
\end{figure}

A detailed study of the relation between dose rate and induced dynamics in the exemplary system \higherrubidium is reported in Fig.~\ref{fig:30Rb_flux}. Here as well as in the following discussion of the dose-rate dependent dynamics in the other systems, for a given system all autocorrelation functions have been fitted with a common coherence factor $\beta$ and shape parameter $\alpha$, and independent decay constants $\Gamma$. Apart from the qualitatively expected monotonic acceleration of dynamics with dose rate, two points may be noted: First, in the regime of comparatively low dose rates, the relation is perfectly linear. Specifically, it is very suggestive that the extrapolation to zero dose rate would indeed correspond to practically frozen dynamics, thus no thermal dynamics accessible by aXPCS are expected at room temperature in this system. On the other hand, at higher dose rates a significant upward deviation from the linear relationship is obvious in \higherrubidium. We will discuss this point in more detail below.

\begin{table}[b]
  \centering
\begin{tabular*}{8cm}{@{\extracolsep{\fill}}cccc}
System&$\Gamma$ (1/s)&$\alpha$&$S_\text{AA}$\\
\hline
B$_2$O$_3$                            &0.187(8) & 0.790(63)&0\\
\lowrubidium                          &0.170(6) & 0.813(56)&0.205\\
\boratecompositionformula{2}{Cs}{98}  &0.255(28)& 0.525(65)&0.370\\
\highrubidium                         &0.081(4) & 0.545(39)&0.692\\
\highcaesium                          &0.163(12)& 0.414(24)&0.832\\
\higherrubidium                       &0.0429(7)& 0.436(12)&0.838
\end{tabular*}
  \caption{Dynamical properties $\Gamma$ and $\alpha$ at structural peak $q_\text{max}$ and evaluated for a dose rate $S=1$~eV/atom and second for the different systems, together with the partial structure factor for the alkali atoms $S_{\text{AA}}$, corresponding to the relative contribution of the alkali atoms to the scattered intensity.}
  \label{systeme_abs}
\end{table}

\begin{figure}
\centering
\includegraphics{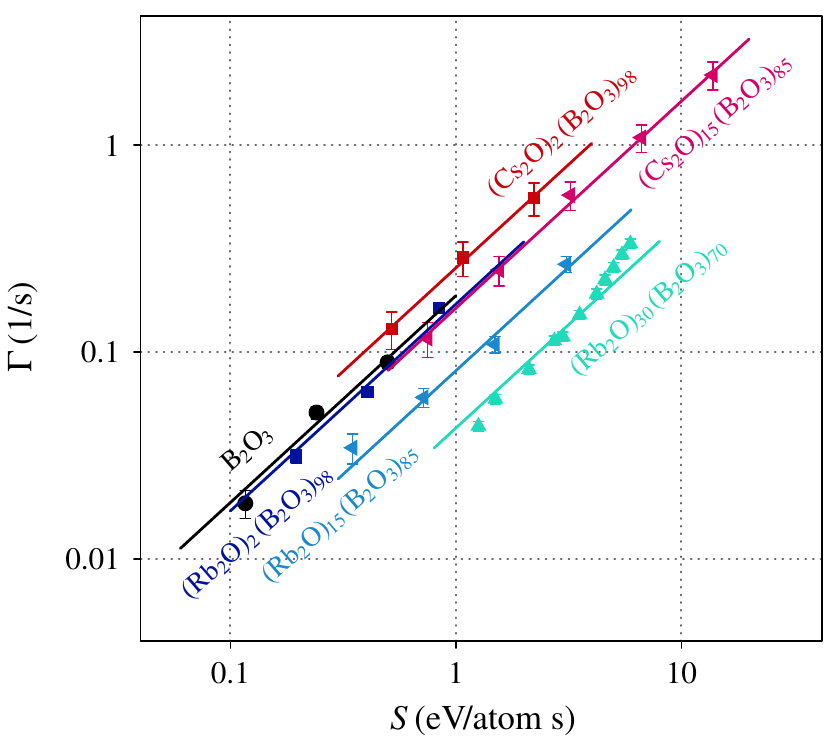}
\caption{Decay constant $\Gamma$ at the respective structure peaks for the different system as function of dose rate together with fits assuming a linear proportionality.}
\label{fig:slopes}
\end{figure}

The linear relationship between dose rate and dynamics is reproduced very satisfactorily also in the other systems as illustrated in Fig.~\ref{fig:slopes}. This figure gives the impression that at a given dose rate the dynamics become slower with increasing alkali content, while the Rb- and Cs-systems for the same alkali content $x$, which a priori would be expected to be chemically quite similar, show noticeable differences with the Cs-based system in general being faster. 

However, it has to be noted that the probed dynamics in aXPCS is to be understood as an average over the dynamics of the constituent elements, which are likely to be different. Further, due to the much larger scattering cross section of Cs compared to Rb, at a given alkali content $x$ the alkali contribution will be larger in the Cs-based than in the Rb-based system. The corresponding values calculated under an assumed absence of short-range order are given in Tab.~\ref{systeme_abs}, together with the decay rates $\Gamma$ evaluated in each system for a dose rate S of 1 eV/atom and second and the shape parameters $\alpha$. The following two hypotheses are now sufficient for explaining the observed behavior:
\begin{itemize}
\item Consistent with the behavior of the glass transition temperature, the structural rearrangements become slower for increasing alkali content $x$, where the introduction of both Rb$_2$O and Cs$_2$O into the matrix is in first approximation equally effective.
\item In any system, the dynamics of rearrangements of the alkali sites is faster than that of the borate matrix.
\end{itemize}
Starting with B$_2$O$_3$, we see that the dynamics of the borate network is described by a shape parameter $\alpha$ close to one, as expected for a strong glass former~\cite{nissjphyscondmat2007}, and as found also previously~\cite{pintori2019relaxation}. Introducing a small amount of alkali oxide ($x=0.02$) does not have a large effect (compare Fig.~\ref{fig:Glass_transition}), so that the values for \lowrubidium stay essentially the same. On the other hand, when introducing the same amount of Cs$_2$O, the weight of the alkali sites in scattered intensity becomes noticeable, leading first to a faster decay rate (as it was assumed that the alkali sites rearrange faster), and further show up in a decrease of the shape parameter $\alpha$ that reflects the scattered intensity being made up of contributions with differing dynamics. At an alkali content of $x=0.15$, the overall slowing of dynamics becomes apparent. Now also the Rb$_2$O-based system has a sizeable contribution from the alkali sites to the scattering and thus a decreased $\alpha$, while in the Cs$_2$O-based system the alkali scattering dominates strongly, leading to faster apparent dynamics with a small shape parameter $\alpha$ that reflects the pertinent autocorrelation function being composed of a fast component with large amplitude on top of a slow component with small amplitude. At $x=0.30$, the deceleration has progressed even further. Note that this slowing-down of structural rearrangements with increased alkali content is quite unrelated to ionic diffusion, as the latter accelerates in the corresponding situation~\cite{berkemeier2005molar}.

\begin{figure*} 
  \includegraphics{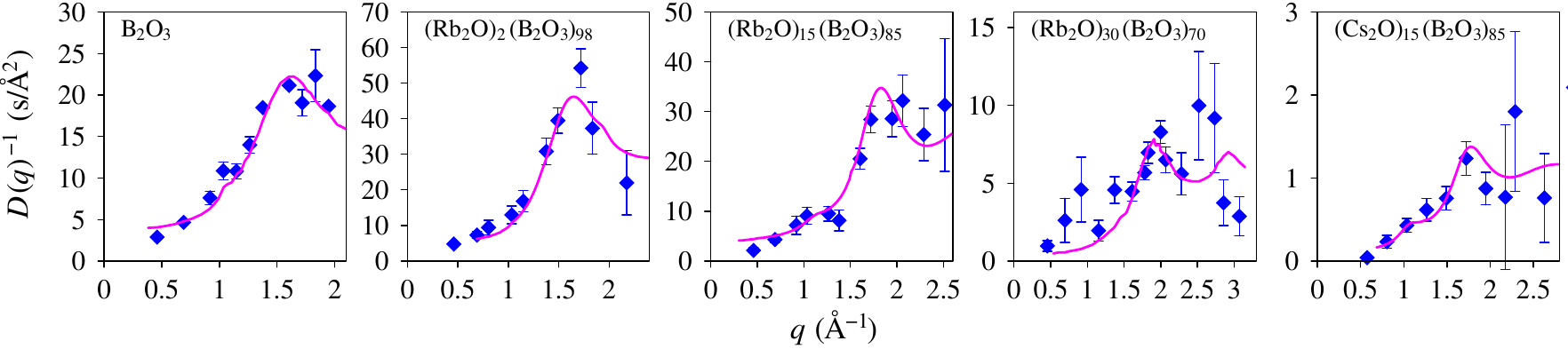}
  \caption{Inverse $q$-dependent diffusion constant $D(q)^{-1}$. Apart from \lowrubidium and \highrubidium, which were measured with a beam attenuated by a factor of $0.483$, all were measured at full beam. The data are compared to the corresponding IBM models (solid lines).}
  \label{fig:Diffusion}
\end{figure*}

\subsection{Spatial aspects of dynamics}
\label{sec:Modelling Beam-induced Dynamics}
We now turn to an explicit microscopic modelling of the structural rearrangements that give rise to the temporal fluctuations in the coherently scattered intensity as studied by aXPCS. Such information is accessible by studying, different from the previous section, how the decay rate $\Gamma(q)$ varies also away from the structural peak at $q_\text{max}$. 

The simplest model for diffusive dynamics is to assume that the atoms independently perform a random walk in the Brownian sense, composed of microscopic steps below the resolution of observation and thus assumed as infinitesimal. In this case, the decay rate has the particularly simple form of 
\be
\Gamma(q)=q^2D_\text{BM},
\ee
where $D_\text{BM}$ is the diffusion constant corresponding to the Brownian motion. Conversely, this relation suggests to reduce our experimentally observed decay constants to $q$-dependent diffusion ``constants''~\cite{segrepre1995}
\be
D(q)=\Gamma(q)/q^2,
\ee
which would be independent of $q$ under the assumption of the atoms performing independent Brownian motion. Measurement series of $\Gamma(q)$ for the small and intermediate $q$-range in all systems considered here are reported in this way in Fig.~\ref{fig:Diffusion}.

Obviously, the $q$-dependent diffusion constants are far from being constant, which rules out the naive assumption of the atoms in our alkali-borate glasses performing independent Brownian motion. In contrast, we observe over all systems a clear increase of $D(q)$ up to about $q=1.8$~\AA$^{-1}$, after which an oscillatory behavior can be conjectured. This is reminiscent of the structure factor $S(q)$, and indeed it has been initially proposed for quasi-elastic neutron scattering on liquids~\cite{DeGennes1959} and later verified also for aXPCS in hard condensed matter~\cite{leitnerjphyscondmat2011} that the simplest modification to incorporate preferential atomic arrangements reflected in a non-constant structure factor $S(q)$ into a given model of independent microscopic motion is to divide the decay constant $\Gamma(q)$ by the structure factor $S(q)$. We term the resulting model
\be
D_\text{IBM}(q)=\frac{D_\text{BM}}{S(q)}
\ee
an interacting Brownian motion (IBM) model. The effect of a slowing of dynamics where the structure factor is high is known as de Gennes-narrowing.

The perfect fit of the IBM model is at odds with the scenario proposed by Pintori et al. for pure borate glass~\cite{pintori2019relaxation}: In their model the effect of an absorbed photon is restricted to a comparatively small number of atoms, but each of these atoms is displaced over a distance on the order of $1/q_\text{max}$. However, these large displacements would lead to a constant $\Gamma(q)$ at $q>q_\text{max}$, as indeed has been found in lead-silicate glasses~\cite{ross2014direct}, or equivalently a strongly decreasing $D(q)$. In contrast, our $q$-resolved data show that the microscopic displacements can be regarded with good accuracy as infinitesimal on \emph{any} probed scale, that is, in the majority they have to be much smaller than the typical atomic distances. 

\section{Microscopic picture of beam-induced dynamics}
In our view, the most plausible model in accordance with the available data is the following: In the majority of cases, the primary event of photoabsorption happens with the emission of an electron from one of the lowest accessible core levels in the system. The resulting deep hole is de-excited by emission of a fluorescence photon, in which case the process of absorption is started anew at somewhat lower energy, or by an Auger electron. In either case, the X-ray photon energy is initially transferred into the electron system in the form of a few electrons in the keV range. These mobile high-energy electrons will then inelastically interact with other electrons, specifically break covalent bonds and lead to structural rearrangements. 

In the initial report on the observation of X-ray beam-induced dynamics, Ruta et al. have called this process radiolysis~\cite{ruta2017hard}. While this term captures the essence of the athermal breaking of bonds, we feel it necessary to point out two main differences to classical cases of radiolysis such as the decomposition of water by alpha radiation: first, here it is not the primary ionizing radiation (in our case the X-ray photons) that break up the majority of the bonds, but rather the emitted Auger and photoelectrons.

Further, the absence of large-scale permanent beam damage as discussed in Sect.~\ref{sec:Structural properties} shows that the breaking of the bonds is only transient: the ejection of electrons from a covalent bond due to the inelastic collisions with the keV-range photo- and Auger electrons leads to a local electron deficiency (and at the same time an electron excess at the sites where the ejected electrons come to rest again). The borate and alkali-borate glasses are insulators with respect to electronic current, thus these localized charges have a finite lifetime during which the electronic structure will decay to its new \emph{local} ground-state subject to the non-neutral charge. As a consequence, on timescales characteristic of phononic dynamics (picoseconds) the atoms will then relax in this modified potential as resulting in the Born-Oppenheimer picture to a new configuration, probably characterized by increased or decreased coordinations determined by the localized charge state. Still, even in insulators electrons have a finite mobility, which is further aided on by disturbances in the electronic structure due to the continuing absorption of photons, so that at even later times the locally neutral charge state with the most advantageous atomic coordinations will be restored. However, the inherent geometrical frustration in these systems, which is the reason for the glassy state being stable at all, implies that there are a number of local configurations more or less equivalent in energy the system can relax into. What leads to structural dynamics on the time-scale of seconds as visible in aXPCS are those events where the final state after restoration of local charge neutrality is different from the intial state. As the rearrangements proceed via athermal relaxations towards new local groundstates due to modified ionic potentials rather than thermal transitions over energy barriers, the large mass differences between chemically equivalent Rb- and Cs-based borate systems will not affect the timescale of dynamics, different from the classical isotope effect of diffusion~\cite{mehrerdiffusion2007}. 

According to our data (see Fig.~\ref{fig:slopes}) and in perfect agreement with Pintori et al.~\cite{pintori2019relaxation}, a timescale of intensity fluctuations at the structural maximum of about one second corresponds to dose rates on the order of \quant{10}{eV/s}. In other words, when \quant{10}{eV} have been deposited per atom, the atomic configuration is essentially different from before. The coincidence of this value with the energy of a few eV necessary for breaking a covalent bond is a strong argument in favor of the correctness of the proposed model, but shows at the same time that the driving of dynamics by the beam is surprisingly efficient, as not all broken environments will relax towards a final state different from the initial state, and also the mobile electrons can lose some of their energy directly to the phononic system as opposed to using it for breaking bonds. Thus it seems that whenever a local rearrangement happens, it is not only the handful atoms involved in the broken and restored bonds that are shifted by distances on the order of half of the typical atomic separations, but that also a large number of atoms in the vicinity are relaxing by smaller distances due to the elastic coupling. This also explains why our IBM model, assuming infinitesimal steps, is so successful in explaining the $q$-dependent data.

We want to note that the proposed mechanism is different from direct knock-on damage, where the primary high-energy particle transfers its energy directly onto an atom. In our case, the maximum possible energy transfer would be about \quant{2}{eV} for the elastic collision of a \quant{13}{keV} photoelectron with a B or O atom, much less than the commonly quoted threshold of \quant{25}{eV}~\cite{kenikphilmag1975}. However, this latter value applies to the formation of a Frenkel pair in a crystal lattice, while for a network glass the threshold could be much lower. Here again it is the agreement with the data obtained by Pintori et al.~\cite{pintori2019relaxation}, who used incident radiation of \quant{8.1}{keV} resulting in correspondingly smaller potential energy transfers, but observe comparable dynamics to our investigation, that allows us to rule out the dominance of this mechanism.

The picture of athermal beam-induced dynamics given above implies that the realized structure on the microscopic scale results essentially from successive relaxations towards local minima in temporally varying potentials. In other words, it would lead to the opposite of a well-relaxed glassy state. We can now ask to which degree this consequence will be accurate as some thermal dynamics will be active even at room temperature. Indeed, the observation of deviations from a perfect linear relationship between dose rate and observed dynamics in \higherrubidium as reported above can be seen as a hint in this direction: possibly some background thermal dynamics are always active in relaxing the local structures resulting from the beam-induced rearrangements. The deviation from linearity at a dose rate of about \quant{3}{eV/atom} and second would then be the point where the rate of beam-induced changes surpasses the rate of thermal relaxation in this most alkali-rich system, leading to less stable configurations and thus a super-linear acceleration of dynamics.

\section{Conclusions}
Studies of rubidium and caesium borate glasses with different ionic concentrations confirm the flux-dependency of the atomic dynamics. No evidence of cumulative beam damage has been found, nor a significant transient effect on the diffraction curves. Hence, the beam induces only dynamical changes within the irradiated volume. 

Further remarkable features are summarized as follows:
\begin{itemize}
\item For a given system, the relation between dose rate and dynamics is linear to a good accuracy. Specifically, inherent thermal dynamics seem to be too slow for being detectable by aXPCS at room temperature. 
\item Consistent with the rise of the glass transition temperature, the rate of dynamics at a given dose rate decrease with increasing alkali content.
\item Comparing the decay rate $\Gamma$ and shape parameter $\alpha$ of chemically equivalent Rb- and Cs-borate systems, it can be inferred that the beam-induced dynamics is faster for the alkali sites than for the borate network.
\item For all systems, the $q$-dependent behavior of the decay rate can be modelled satisfactorily by assuming the atoms to perform Brownian motion composed of infinitesimal jumps, subject to de Gennes narrowing.
\end{itemize}
Our studies serve to underline the point recently raised~\cite{leitner2015acceleration} and subsequently experimentally verified~\cite{ruta2017hard}: the absence of beam damage in the sense of a permanent modification of the sample's structure in a scattering experiment is a necessary, but no sufficient condition for assuming the \emph{dynamical properties} to be representative of their inherent values. Concerning experiments at next generation ultralow emittance synchrotrons or free-electron laser sources using high intense coherent beams that are based on fs pulses, the beam-induced effect will become more and more relevant and therefore should be studied thoroughly to prevent misinterpretations of the experimental results. On the other hand, it is obvious that this possibility to directly study the dynamics of how hard X-ray photons below the knock-on energy threshold can rearrange matter on the atomic scale for the first time will lead to an improved understanding of the processes that do lead to permanent irradiation damage in materials, and thus are of utmost technological relevance, for instance for future fusion reactors.

\section{Acknowledgments}
This work was funded by the Austrian Science Fund (FWF): P28232-N36.

We are greatful to Dr Michael Sprung (beamline scientist at DESY) for his help performing the aXPCS experiments at the beamline P10 at DESY. We want to thank Werner Artner for the support of the WAXS measurements at the Technical University of Vienna.

\bibliography{abkuerz,literatur}
\onecolumngrid
\end{document}